\documentclass[review]{vgtc}

\graphicspath{{figures/}{pictures/}{images/}{./}} 

\usepackage{booktabs}                  
\usepackage{multirow}                  
\usepackage{threeparttable}            
\usepackage{colortbl}                  
\usepackage{capt-of}                   
\usepackage{balance}                   

\usepackage[T1]{fontenc}
\usepackage{amsmath,amssymb}
\usepackage{newtxtext,newtxmath}

\definecolor{ResStreamGreen}{HTML}{EAF4E4}

\DeclareMathOperator{\Render}{Render}
\DeclareMathOperator{\Compose}{Compose}
\DeclareMathOperator{\sg}{sg}

\onlineid{}

\vgtccategory{Research}

\vgtcinsertpkg

\title{ResStreamGS: Residual-Sensitive Dual-Stream Gaussian Splatting for \\ Mobile Volumetric Video Streaming}

\author{Haoran Liu, Di Wu, Wang Luo, Hengyuan Na, Miao Hu, and Guocong Quan}
\hypersetup{
  pdftitle={ResStreamGS: Residual-Sensitive Dual-Stream Gaussian Splatting for Mobile Volumetric Video Streaming},
  pdfauthor={Haoran Liu, Di Wu, Wang Luo, Hengyuan Na, Miao Hu, and Guocong Quan}
}

\teaser{
  \centering
  \begin{minipage}[c]{0.418\linewidth}
    \centering
    \includegraphics[width=\linewidth,trim=0pt 0 567pt 0,clip]{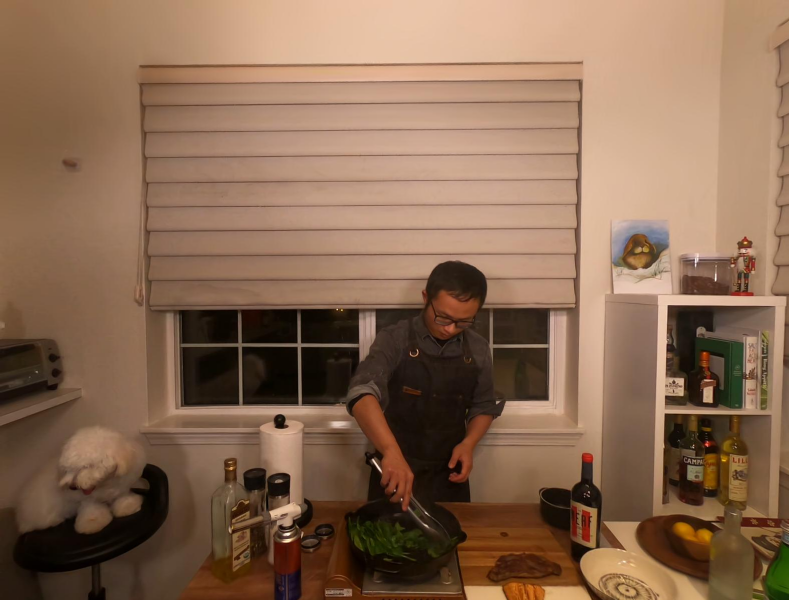}
  \end{minipage}\hspace{0.5pt}%
  \begin{minipage}[c]{0.418\linewidth}
    \centering
    \includegraphics[width=\linewidth,trim=394pt 0 173pt 0,clip]{teaser_resstreamgs.pdf}
  \end{minipage}\hspace{0.5pt}%
  \begin{minipage}[c]{0.160\linewidth}
    \centering
    \includegraphics[width=\linewidth]{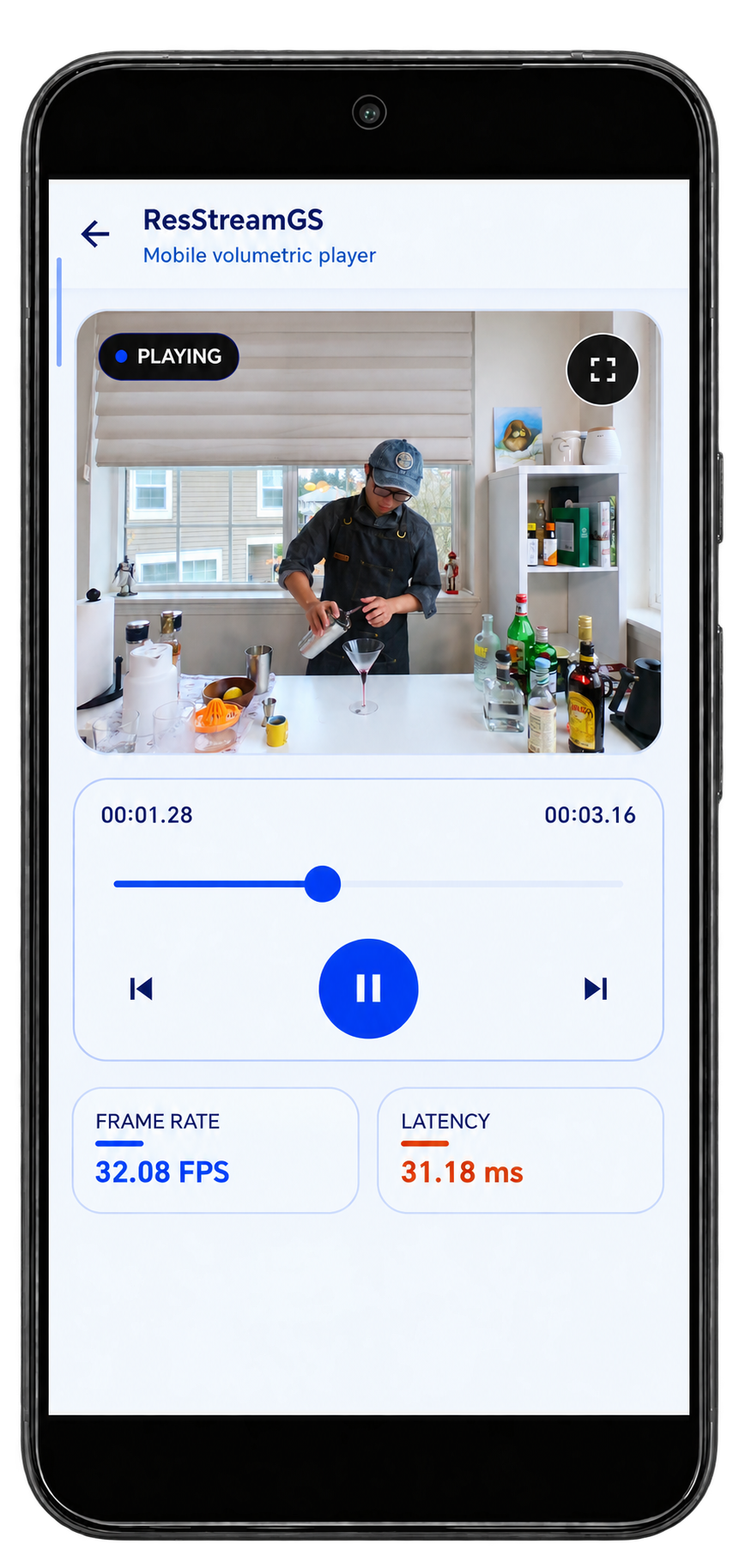}
  \end{minipage}
  \vspace{-0.4cm}
  \caption{Representative full-scene volumetric video scenarios and our mobile playback.}
  \label{fig:teaser}
}

\abstract{
Real-time mobile volumetric video enables immersive free-viewpoint experiences, but delivering high-quality 3D content under constrained bandwidth and computation remains challenging. Recent 3D Gaussian Splatting (3DGS) systems, such as V$^3$, enable efficient mobile streaming by encoding time-varying Gaussian attributes with commodity video codecs and hardware decoding. However, existing pipelines treat all Gaussians as frame-varying signals, which is poorly suited to full-scene content where large portions remain temporally stable. Our empirical study shows that frame-wise adaptation and video coding of all Gaussians cause quality degradation in complex full-scene content, particularly in temporally stable regions. To address this issue, we propose \textit{ResStreamGS}, a residual-sensitive dual-stream framework that processes Gaussians according to their temporal behavior. ResStreamGS identifies residual-sensitive primitives through lightweight probe adaptation and reuses stable Gaussians within each Group of Pictures (GOP). Stable Gaussians are transmitted once per GOP, while only residual-sensitive Gaussians remain in the frame-wise hardware-decodable video stream. Experiments on complex full-scene volumetric videos show that ResStreamGS improves post-codec PSNR by 7.51--8.35~dB over a controlled V$^3$-style all-Gaussians baseline, reduces amortized payload by up to 56.5\%, and achieves 30.883~FPS on a Qualcomm Snapdragon 778G platform.
} 


\begin{document}


\firstsection{Introduction}

\maketitle

Volumetric video enables users to freely explore dynamic 3D scenes from arbitrary viewpoints and supports immersive applications such as virtual reality, mixed reality, and remote telepresence. Bringing these experiences to commodity mobile devices remains challenging because a practical system must maintain high visual quality while reducing transmission, decoding, memory, and rendering costs. 3D Gaussian Splatting (3DGS) provides a promising representation for this setting because its explicit point-based formulation supports high-quality reconstruction and efficient rasterization~\cite{kerbl2023gaussian}. To further reduce transmission and decoding costs, recent systems organize time-varying Gaussian attributes into 2D video-like representations and compress them more effectively with commodity hardware codecs such as H.264 and HEVC~\cite{wang2024v3,lee2025codecgs,mpeg2026gsc}.

Existing Gaussian-video pipelines encode the complete Gaussian representation as a frame-varying signal. This design is effective for human-centric or foreground-dominant content, where many important primitives change together with the moving subject. Complex full-scene volumetric videos have a different temporal structure. Dynamic regions occupy only part of the scene, while large portions of the geometry and appearance remain persistent over time. Treating all Gaussians in such scenes as frame-varying signals may therefore introduce unnecessary processing and distortion.

\begin{figure*}[t]
  \vspace{-0.4em}
  \centering
  \begin{minipage}[t]{0.48\textwidth}
    \centering
    \begin{minipage}[c][104pt][c]{\linewidth}
      \centering
      \includegraphics{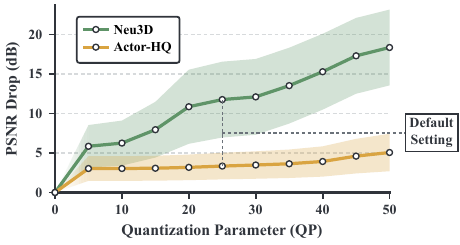}
    \end{minipage}
    \begin{minipage}[t][4.5\baselineskip][t]{\linewidth}
      \captionof{figure}{PSNR degradation of V$^3$ across H.264 quantization parameters on Neu3D and Actor-HQ. Neu3D represents full-scene content with persistent backgrounds, Actor-HQ contains foreground human subjects.}
      \label{fig:v3_quality_drop}
    \end{minipage}
  \end{minipage}\hfill
  \begin{minipage}[t]{0.475\textwidth}
    \centering
    \begin{minipage}[c][104pt][c]{\linewidth}
      \centering
      \raisebox{9pt}{%
      \begin{minipage}[t]{0.485\linewidth}
        \centering
        \includegraphics[width=\linewidth,height=88pt,trim=0 5pt 0 0,clip]{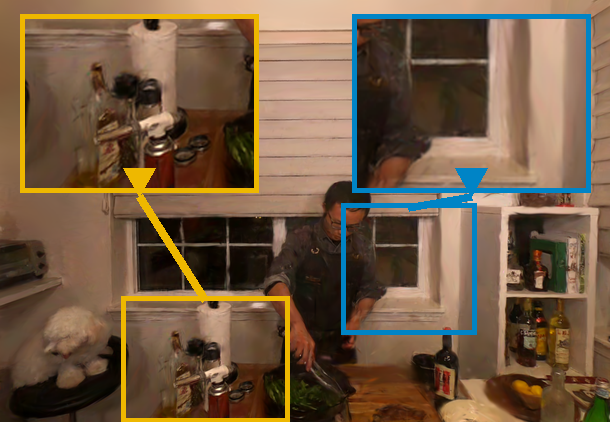}\\[1.125em]
        \footnotesize\textbf{(a) V$^3$}
      \end{minipage}%
      }\hfill
      \raisebox{9pt}{%
      \begin{minipage}[t]{0.485\linewidth}
        \centering
        \includegraphics[width=\linewidth,height=88pt,trim=0 5pt 0 0,clip]{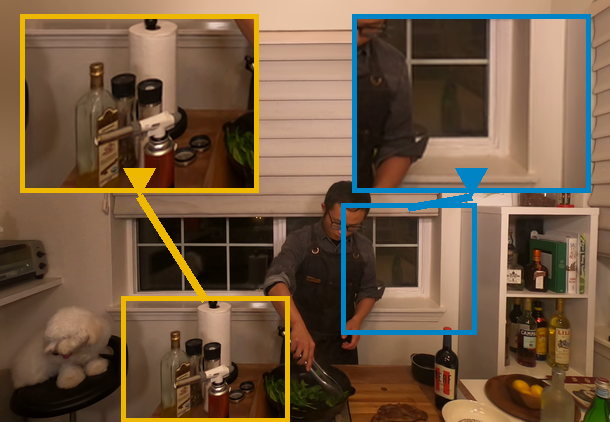}\\[1.125em]
        \footnotesize\textbf{(b) ResStreamGS}
      \end{minipage}%
      }
    \end{minipage}
    \begin{minipage}[t][4.5\baselineskip][t]{\linewidth}
      \captionof{figure}{Qualitative comparison on \textit{cook\_spinach}. V$^3$ introduces visible distortions in scene structures. ResStreamGS preserves these regions with fewer artifacts and better appearance details.}
      \label{fig:qualitative_comparison}
    \end{minipage}
  \end{minipage}
  \vspace{-0.8em}
\end{figure*}

To better understand this limitation, we conduct an empirical study by applying the same V$^3$-style H.264 Gaussian-video pipeline to both foreground-focused and complex full-scene content. As shown in \cref{fig:v3_quality_drop}, Neu3D consistently exhibits a much steeper PSNR decline across increasing QP values than Actors-HQ; under the default setting, the drops are 11.7~dB and 3.4~dB, respectively. More importantly, the qualitative results in \cref{fig:qualitative_comparison} show that the errors are not limited to rapidly moving regions. Scene structures that should remain largely stable can also become visibly corrupted. These results suggest that the all-Gaussians formulation is poorly matched to the temporal structure of complex full-scene content.

These results further motivate us to examine why full-scene content suffers more degradation. Our analysis in Sec.~\ref{sec:motivation} shows that the key issue lies in a mismatch between the temporal behavior of Gaussian primitives and the granularity at which they are processed. Existing pipelines repeatedly adapt and encode every primitive at the frame level, even though many primitives change little within a short frame group. As a result, stable primitives are exposed to optimization drift and codec distortion without providing benefits in temporal adaptability. This observation suggests that Gaussian primitives should not be uniformly processed at the same temporal granularity. Stable primitives can instead be reused across frames, while only primitives sensitive to temporal residuals need to remain in the frame-wise adaptation and video-coding path.

Motivated by these findings, we propose \textbf{ResStreamGS}, a residual-sensitive dual-stream framework for efficient full-scene volumetric video streaming. ResStreamGS first uses lightweight probe adaptation to identify primitives that are sensitive to temporal changes. It then selects the smallest residual-sensitive subset that preserves the reconstruction quality of the composed scene. The remaining primitives are treated as stable references and reuse their keyframe attributes within the Group of Pictures (GOP). This partition produces two synchronized streams with different temporal granularities. Stable Gaussians are transmitted once per GOP and cached on the mobile device, while residual-sensitive Gaussians remain in a compact frame-wise Gaussian-video stream that can be decoded with commodity video hardware. In this way, ResStreamGS retains the advantages of hardware video decoding while removing persistent scene state from repeated frame-wise processing.

Experiments on full-scene volumetric videos demonstrate that this design substantially improves both reconstruction fidelity and streaming efficiency. Under the same Gaussian-video coding framework, ResStreamGS improves post-codec PSNR by 7.51--8.35~dB over the controlled V$^3$ all-Gaussians baseline while reducing the amortized payload by up to 56.5\%. Our mobile implementation achieves 30.883~FPS on a Qualcomm Snapdragon 778G mobile platform, showing that high-quality full-scene volumetric video can be played in real time on commodity mobile hardware.

\textbf{Our Contributions:} (1) \underline{\textbf{\textit{New Perspective.}}} We identify a temporal mismatch in Gaussian-video streaming, where primitives with different temporal behaviors are uniformly processed at the frame level. This motivates processing Gaussians according to their temporal sensitivity.
(2) \underline{\textbf{\textit{Novel Framework.}}} We propose ResStreamGS, a residual-sensitive dual-stream framework that separates stable Gaussians for GOP-level reuse from residual-sensitive Gaussians for frame-level adaptation, with quality-aware selection to preserve reconstruction quality.
(3) \underline{\textbf{\textit{Superior Performance.}}} ResStreamGS substantially improves post-codec reconstruction quality while reducing the amortized streaming payload by up to 56.5\%, and consistently achieves 30.883~FPS on a Qualcomm Snapdragon 778G mobile platform for smooth real-time full-scene volumetric video playback in practical mobile deployment scenarios.
\section{Related Work}

\subsection{Dynamic Scene Reconstruction via 3DGS}

Dynamic novel view synthesis (NVS) reconstructs time-varying 3D scenes from videos for novel-view rendering. Early methods extend NeRF with deformation fields, scene flow, or time conditioning to model dynamic content \cite{mildenhall2021nerf,pumarola2021dnerf,park2021nerfies,li2021nsff,tretschk2021nonrigid,gao2021dynamicview}. Later explicit and factorized representations improve efficiency using voxel grids, multi-plane decompositions, or tensorized 4D structures \cite{fang2022tineuvox,wang2022fourier,cao2023hexplane,fridovichkeil2023kplanes,shao2023tensor4d}, but still rely on costly volumetric sampling or neural decoding.

3D Gaussian Splatting (3DGS) instead adopts an explicit point-based representation with rasterization-oriented rendering \cite{kerbl2023gaussian}. Dynamic 3DGS methods further model temporal changes through deformation fields, Gaussian tracking, frame-wise updates, or pose-dependent Gaussian maps \cite{yang2024deformable3dgs,luiten2024dynamic3dgaussians,li2024animatable,hu2024gaussianavatar,zheng2024gpsgaussian,zhou2025gps,zielonka2023drivable,li2024spacetimegs}. Although these methods achieve high reconstruction quality and rendering speed, their large per-frame storage and limited compatibility with commodity video codecs make direct deployment for mobile volumetric video streaming challenging \cite{yang2024deformable3dgs,luiten2024dynamic3dgaussians,li2024spacetimegs}.

\begin{figure*}[t]
  \centering
  \includegraphics[width=0.94\textwidth]{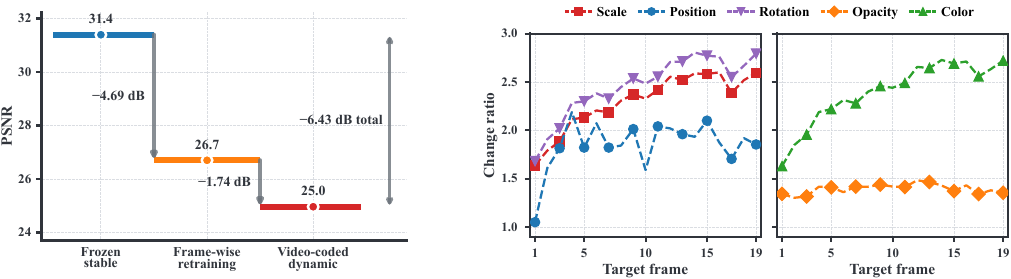}\par
  \vspace{-0.2em}
  \makebox[0.45\textwidth][c]{\footnotesize\textbf{(a) Stage-wise PSNR loss}}\hspace{0.04\textwidth}%
  \makebox[0.45\textwidth][c]{\footnotesize\textbf{(b) Attribute perturbation}}\par
  \caption{(a) Stage-wise PSNR degradation in nominally stable regions: the frozen configuration defines the 31.4-dB reference, while frame-wise retraining and video coding incur losses of 4.69 and 1.74~dB, respectively, totaling 6.43~dB. (b) Gaussian attribute change ratios across target frames within the GOP.}
  \label{fig:intro_diagnostic}
\end{figure*}

\subsection{Efficient and Compact 3D Gaussian Representations}

Despite efficient rasterization, 3DGS introduces substantial storage overhead because each Gaussian stores geometry, opacity, and view-dependent appearance attributes \cite{kerbl2023gaussian}. Existing compact 3DGS methods mainly reduce this cost through opacity- or contribution-aware pruning, vector quantization, entropy coding, codebook-based clustering, and context-adaptive prediction \cite{lee2024compact3dgs,morgenstern2023compact,niedermayr2024compressed,fan2024lightgaussian,chen2024hac,lee2025hierarchical}. These techniques show that Gaussian representations contain spatial and attribute redundancy and can be compressed without severely degrading rendering quality. However, they are designed for static scenes, where the representation is decoded once and reused across viewpoints. Dynamic Gaussian sequences further amplify the storage burden as geometry and appearance evolve over time. Recent methods therefore exploit temporal redundancy through keyframes, residual modeling, motion-aware updates, and memory-efficient 4D parameterizations \cite{lee2024c3dgs,zhang2025mega,dai2025fourDGV,yuan2025fourDGS1K}. Streaming-oriented approaches employ adaptive layering and saliency-aware tiling to reduce transmission cost \cite{gong2025adaptive3dgs,gong2025saliency3dgs}. Nevertheless, most methods optimize how the complete Gaussian sequence is compressed, without distinguishing primitives that remain stable over time. Our work instead separates stable Gaussians from residual-sensitive ones before dynamic encoding, avoiding repeated lossy compression of stable structures and reserving the dynamic stream for time-varying content.

\subsection{Mobile Volumetric Video Streaming}

Mobile volumetric video systems must jointly reduce network bandwidth, decoding overhead, and rendering cost on mobile devices. Early systems based on point clouds or meshes employ visibility-aware scheduling, adaptive transmission, view-dependent content selection, and edge-assisted processing to support interactive playback under constrained resources \cite{han2020vivo,zhang2022yuzu,guan2023metastream}. Recent 3DGS systems extend these ideas by adapting transmitted content, quality, or bitrate according to network conditions and user viewpoints \cite{yin2024fsvfg,sun2025lts,liao2025static3dgs,gong2025adaptive3dgs,gong2025saliency3dgs,cags2026,gsstream2026}. SwinGS~\cite{swings2024} further adopts a sliding-window design to limit memory consumption and support longer volumetric video sequences. Video-coded Gaussian streaming takes a complementary direction by exploiting commodity video hardware. V$^3$~\cite{wang2024v3} maps dynamic Gaussian attributes into 2D Gaussian videos, enabling H.264 compression, hardware decoding, and shader-based rendering on mobile devices. Vega~\cite{kim2025vega} further improves full-scene mobile streaming through object-level selective computation and view-adaptive scheduling, while MobileGS and SEELE optimize client-side Gaussian processing and rasterization \cite{mobilegs2026,seele2026}. Despite these advances, existing video-coded pipelines still treat all Gaussians as temporally varying signals and repeatedly pass them through frame-wise compression. This can unnecessarily expose stable background primitives to quantization noise and optimization-induced attribute drift. Our work instead isolates stable Gaussians before encoding, preserving them across frames while retaining compact dynamic streams for real-time mobile playback.
\section{Preliminary}
\label{sec:preliminary}

3D Gaussian Splatting (3DGS)~\cite{kerbl2023gaussian} represents a scene as a set of anisotropic Gaussian primitives $G=\{g_i\}_{i=1}^{N}$. Each primitive is parameterized as
\begin{equation}
    g_i =
    \left(
    \mathbf{p}_i,\mathbf{q}_i,\mathbf{s}_i,
    \alpha_i,\mathbf{c}_i
    \right),
\end{equation}
with $\mathbf{p}_i$, $\mathbf{q}_i$, $\mathbf{s}_i$, $\alpha_i$, and $\mathbf{c}_i$ describing its position, rotation, scale, opacity, and view-dependent color, respectively. The rotation and scale determine the spatial extent of a Gaussian through its covariance matrix. Under a viewing transformation $\mathbf{W}$, the covariance is projected onto the image plane as
\begin{equation}
    \mathbf{\Sigma}_i =
    \mathbf{R}_i\mathbf{S}_i\mathbf{S}_i^{T}\mathbf{R}_i^{T},
    \qquad
    \mathbf{\Sigma}'_i =
    \mathbf{J}\mathbf{W}\mathbf{\Sigma}_i
    \mathbf{W}^{T}\mathbf{J}^{T},
\end{equation}
in which $\mathbf{R}_i$ and $\mathbf{S}_i$ are the rotation and scaling matrices, $\mathbf{J}$ is the Jacobian of the projective transformation, and $\mathbf{\Sigma}'_i$ gives the projected covariance. During rendering, the Gaussians overlapping a pixel are depth-sorted and accumulated through front-to-back alpha blending:
\begin{equation}
    \mathbf{C} =
    \sum_{i\in\mathcal{N}}
    \mathbf{c}_i\alpha'_i
    \prod_{j=1}^{i-1}(1-\alpha'_j),
\end{equation}
where $\mathcal{N}$ contains the ordered Gaussians contributing to the pixel and $\alpha'_i$ is the corresponding projected opacity. This explicit representation enables efficient rasterization while maintaining high-quality novel-view rendering. For dynamic scenes, a volumetric video can therefore be viewed as a temporal sequence of these Gaussian attributes, whose values evolve from frame to frame.

Building on this explicit representation, V$^3$~\cite{wang2024v3} organizes dynamic 3DGS into frame groups for volumetric video streaming. The first frame of each group is reconstructed as a keyframe, and subsequent frames are optimized sequentially from the preceding Gaussian representation. Motion is first estimated for the Gaussian primitives, followed by fine-tuning of the remaining attributes to improve temporal consistency across frames. By maintaining a consistent set of primitives within each frame group, the attributes of the same Gaussian can be associated across consecutive frames rather than reconstructed independently at every time step. After training, the per-frame Gaussian attributes are packed into 2D attribute images for video coding. V$^3$ applies Morton sorting before packing so that spatially neighboring Gaussians are placed at nearby image locations, improving the spatial coherence of the resulting attribute maps. The same ordering is retained across attribute images and consecutive frames, allowing corresponding pixel locations to represent the same Gaussian primitive over time. As a result, the evolution of Gaussian attributes is converted into temporally coherent image sequences that can be compressed using conventional video codecs such as H.264. On mobile devices, hardware video decoders recover the attribute maps for standard 3DGS rendering.

\section{Motivation}
\label{sec:motivation}

In this section, we investigate why existing all-Gaussians video coding is poorly suited to complex full-scene volumetric video. This leads to our first research question.

\subsection{RQ1. Is All-Gaussians Video Coding Effective for Full-Scene Content?}
\label{sec:motivation:failure}

\noindent\textbf{Effectiveness Analysis.}
Existing Gaussian-video pipelines treat the complete Gaussian representation as a frame-varying signal, which is well suited to human-centric content where many important primitives change together with the moving subject. Full-scene sequences such as \textit{cook\_spinach} and \textit{coffee\_martini}, in contrast, contain localized dynamic regions together with large persistent backgrounds and room-scale structures~\cite{li2022neural3dvideo}. To examine whether the same coding paradigm remains effective under this different temporal structure, we apply the V$^3$-style H.264 pipeline to both foreground-focused Actors-HQ and full-scene Neu3D across different quantization settings. As shown in \cref{fig:v3_quality_drop}, Neu3D exhibits substantially faster PSNR degradation as QP increases, whereas Actors-HQ remains comparatively stable. Under the default setting, the corresponding drops are 11.7~dB and 3.4~dB, respectively, making the degradation on full-scene content more than three times larger. This consistent gap indicates that all-Gaussians video coding is substantially less effective for complex full-scene content.

\noindent\textbf{Error Analysis.}
To understand where this degradation occurs, we further examine the reconstructed scenes after Gaussian-video coding. As shown in \cref{fig:qualitative_comparison}, the errors are not limited to rapidly moving objects or regions with strong temporal changes. Scene structures that should remain largely stable can also become visibly corrupted. This suggests that the degradation cannot be explained only by the difficulty of representing dynamic content. Instead, the frame-wise pipeline may unnecessarily perturb scene content that could otherwise remain persistent.

\noindent\textbf{Finding of RQ1.}
All-Gaussians video coding is substantially less effective for complex full-scene content than for foreground-dominant content, with considerable degradation even in scene structures that should remain temporally stable.

This finding raises our second research question.

\subsection{RQ2. Why Does All-Gaussians Video Coding Degrade Stable Gaussians?}
\label{sec:motivation:diagnosis}

\noindent\textbf{Stage-Wise Analysis.}
To identify the source of this degradation, we analyze nominally stable regions in \textit{cook\_spinach} through successive stages of the reconstruction and coding pipeline. As shown in \cref{fig:intro_diagnostic}(a), using the frozen Gaussians as the reference, frame-wise retraining alone introduces a 4.69~dB PSNR drop, while subsequent video coding causes another 1.74~dB drop, resulting in a total degradation of 6.43~dB. This decomposition shows that stable scene content is affected by both frame-wise optimization and lossy video coding rather than by codec distortion alone.

\noindent\textbf{Temporal Mismatch Analysis.}
The attribute-level analysis in \cref{fig:intro_diagnostic}(b) further shows that frame-wise retraining perturbs multiple Gaussian attributes even for primitives whose underlying scene content changes little. These slightly changed attributes are then repeatedly quantized and reconstructed as temporal video signals, introducing additional distortion. This problem is particularly severe for full-scene content because a large portion of its Gaussians correspond to persistent geometry and appearance within a short GOP. Frame-wise processing provides little benefit for these primitives, yet the all-Gaussians pipeline still repeatedly adapts and encodes them. Importantly, temporal stability does not correspond directly to a semantic foreground-background partition: some spatially stationary primitives still require frame-dependent updates, while some primitives near dynamic regions remain sufficiently stable for reuse. The key distinction should therefore be whether each primitive is sensitive to temporal changes that affect reconstruction quality.

\noindent\textbf{Finding of RQ2.}
The degradation arises from a mismatch between primitive-level temporal behavior and frame-level processing. Uniformly adapting and encoding all Gaussians unnecessarily exposes reusable stable primitives to both optimization drift and codec distortion during repeated frame-wise processing stages.

\noindent\textbf{Insight for ResStreamGS.}
Gaussian primitives should be processed according to their temporal sensitivity rather than uniformly at the frame level. Stable primitives can be represented once and reused within a GOP, while only residual-sensitive primitives need to remain in the frame-wise adaptation and video-coding path.

\section{ResStreamGS}
\label{sec:method}
\label{sec:dual_stream_optimization}
\label{sec:training}

\begin{figure*}[t]
  \vspace{-0.6em}
  \centering
  \includegraphics[width=0.94\textwidth]{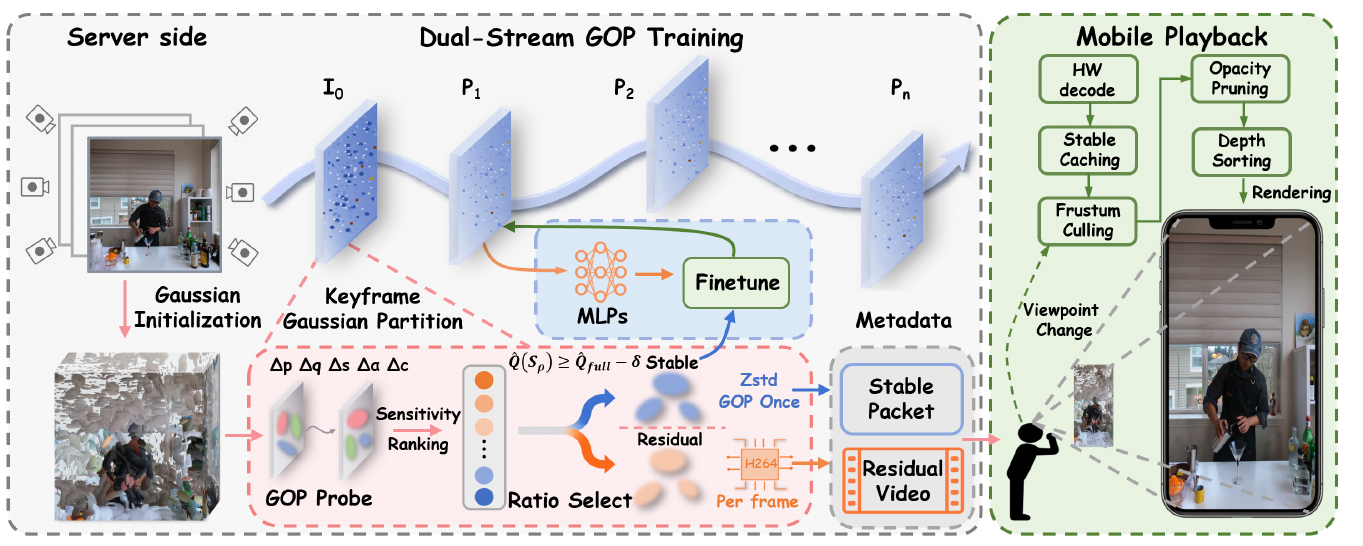}
  \caption{Overview of ResStreamGS. Each GOP is partitioned into group-level stable Gaussians and frame-wise residual-sensitive Gaussians. Blue elements indicate the stable Gaussians stream path, and orange elements indicate the residual-sensitive path.}
  \label{fig:overview}
\end{figure*}

Motivated by the above insight, we propose \textbf{ResStreamGS}, an end-to-end residual-sensitive dual-stream framework for efficient mobile 3DGS streaming, as illustrated in Fig.~\ref{fig:overview}. ResStreamGS organizes Gaussians according to their temporal sensitivity into a group-level stable stream and a frame-level residual-sensitive stream. The framework consists of server-side dual-stream GOP processing and mobile-side composition and rendering.

On the server side, ResStreamGS initializes the Gaussian representation from the keyframe and performs lightweight GOP probing to profile multi-attribute residual sensitivity. It then ranks the primitives according to their sensitivity and applies quality-aware ratio selection to partition them into stable and residual-sensitive subsets. Stable Gaussians are retained at the GOP level and reused across its frames, while residual-sensitive Gaussians undergo frame-wise motion estimation and fine-tuning to capture temporal variations. After optimization, the two subsets are organized into different transmission forms: stable Gaussians are encoded as a stable packet transmitted once per GOP, whereas residual-sensitive Gaussians are arranged into a compact frame-wise Gaussian video compatible with hardware-supported codecs for efficient transmission.

On the mobile side, the stable packet is received once at the beginning of each GOP and cached in device memory for reuse across subsequent frames. In parallel, the residual-sensitive Gaussian video is decoded frame by frame using commodity hardware video decoding. For each frame, the decoded residual-sensitive Gaussians are combined with the cached stable Gaussians to reconstruct the complete Gaussian representation for the current scene state. Given the current user viewpoint, ResStreamGS further applies view-frustum culling to discard invisible primitives, followed by opacity-based pruning to remove Gaussians with negligible rendering contributions. The remaining Gaussians are then depth-sorted and rendered to produce the final view. In this way, the mobile client avoids repeatedly decoding and storing persistent scene content while retaining frame-wise updates for temporally sensitive regions.

\subsection{Frame-Group Reference and Probing}

For each frame group (GOP) $\mathcal{F}=\{t_0,\ldots,t_0+K-1\}$, ResStreamGS takes the keyframe Gaussian set $G_{t_0}$ as its group-level reference. The keyframe establishes primitive correspondence and provides the initial values for all Gaussian attributes. The last frame of $\mathcal{F}$ serves as the probe target. A lightweight probe temporarily adapts the keyframe representation to this target using the calibrated training views of the probe target frame. During adaptation, the probe records per-primitive attribute changes. These observations support residual-sensitivity ranking and the subsequent quality-aware selection of the residual-sensitive subset.

\subsection{Residual-Sensitivity Profiling}

Position alone is insufficient to characterize residual sensitivity because other attributes may change a primitive's projected contribution even when its center remains nearly fixed. We profile per-primitive changes over the attribute set $\mathcal{A}=\{\mathbf{p},\mathbf{q},\mathbf{s},\alpha,\mathbf{c}\}$, comprising position, quaternion rotation, scale, opacity, and DC color.

Let $a_i^0$ denote the keyframe value of attribute $a$ for primitive $i$, and let $a_i^u$ denote its value at a sampled state $u$ along the probe adaptation trajectory after burn-in. We use Euclidean distance for position, scale, and color, quaternion angular distance for rotation, and absolute difference for opacity. For each attribute, we compute the trajectory variance $v_i^a=\operatorname{Var}_{u\in\mathcal{W}}[d_a(a_i^u,a_i^0)]$, where $\mathcal{W}$ contains the sampled states after burn-in. These attribute-specific variances describe the instability observed while adapting the keyframe representation to the probe target.

Before aggregation, we normalize each attribute-specific statistic independently across the primitives in the frame group using
\begin{equation}
    \operatorname{RNorm}_a(x_i^a)
    =
    \operatorname{clip}
    \left(
    \frac{x_i^a-\operatorname{median}(\mathbf{x}^a)}
         {\operatorname{MAD}(\mathbf{x}^a)+\epsilon},
    0,z_{\max}
    \right).
\end{equation}
Here, $\mathbf{x}^a$ collects $x_i^a$ over all primitives in the frame group, $\operatorname{MAD}$ denotes the median absolute deviation, $\epsilon$ is a small constant for numerical stability, and $z_{\max}$ is the clipping threshold. The combined residual-sensitivity score is
\begin{equation}
    r_i
    =
    \sum_{a\in\mathcal{A}}
    w_a\,\operatorname{RNorm}_a(v_i^a),
\end{equation}
with equal attribute weights in the current implementation.

We rank primitives by the multi-attribute instability score $r_i$ obtained from probe adaptation. Lower scores indicate greater suitability for group-level reuse, whereas higher-scoring primitives are prioritized for frame-wise adaptation. The ranking is based on reconstruction-related attribute changes rather than semantic object labels, sensitive background structures may also receive high scores when their Gaussian attributes vary noticeably over time.

\subsection{Quality-Aware Stream Partitioning}

Given the sensitivity ranking, we construct nested candidate index sets. For a residual-sensitive ratio $\rho$, referred to as the dynamic ratio in the experiments, the corresponding candidate index set is
\begin{equation}
    S_{\rho}
    =
    \operatorname{Top}_{\lceil\rho N\rceil}
    \left(\{r_i\}_{i=1}^{N}\right).
\end{equation}
Here, $N=|G_{t_0}|$ is the number of Gaussians in the keyframe representation. All candidate index sets are derived from the same ranking, so $S_{\rho_1}\subset S_{\rho_2}$ whenever $\rho_1<\rho_2$. Changing $\rho$ adjusts the number of selected primitives without changing the underlying ranking. The scores are ordered in descending order, so each Top-$K$ set retains the primitives with the highest residual-sensitivity scores.

For each candidate index set, primitives indexed by $S_{\rho}$ use their probe-adapted attributes at the probe target frame, while the remaining primitives retain their keyframe attributes. Both subsets are rendered together by the Gaussian renderer. We denote by $\widehat{Q}(S_{\rho})$ the PSNR computed from the MSE pooled across the calibrated training views. The reference quality $\widehat{Q}_{\mathrm{full}}$ is obtained by assigning probe-adapted attributes to all primitives. Each candidate index set is therefore evaluated on the complete composed scene.

We select the smallest candidate ratio whose composed-scene quality remains within an allowed degradation $\delta$ from the full-probe reference:
\begin{equation}
    \rho^*
    =
    \min\left\{
    \rho\in\Omega\;\middle|\;
    \widehat{Q}(S_{\rho})
    \ge
    \widehat{Q}_{\mathrm{full}}-\delta
    \right\},
\end{equation}
where $\Omega$ is the finite set of candidate ratios evaluated by the search and $\delta$ is the permitted probe-quality degradation. If no candidate satisfies the constraint, the largest ratio in $\Omega$ is used.

This rule directly minimizes the scope of frame-wise adaptation among the candidate partitions that preserve the required probe quality. It does not use post-codec PSNR or the fixed-ratio sweep reported in the evaluation; codec-level payload is measured separately after compression rather than during selection.

The selected index set $S_{\rho^*}$ defines $G^{r}=\{g_i\mid i\in S_{\rho^*}\}$; the remaining keyframe Gaussians form $G^{s}=G_{t_0}\setminus G^{r}$. This partition specifies which primitives reuse their keyframe attributes across the GOP and which retain frame-dependent attributes.

\subsection{Selective Frame-Wise Optimization}

The stable and residual-sensitive sets continue to form a single scene after partitioning. For frame $t\in\mathcal{F}$, the rendered image is
\begin{equation}
    \widehat{\mathbf{I}}_t
    =
    \Render
    \left(
    \sg(G^{s})
    \cup
    G^{r}_t
    \right),
\end{equation}
where $\Render$ denotes the differentiable Gaussian renderer and $\sg(\cdot)$ denotes the stop-gradient operation. Stopping gradients through $G^{s}$ freezes its keyframe attributes without removing the stable primitives from joint rendering and compositing of the complete scene. Only the parameters of $G^{r}_t$ are optimized during this stage.

For each non-keyframe, a residual translation field updates the primitive centers relative to their preceding-frame positions. The translated state is then fine-tuned to adjust geometry and appearance attributes. The stable set remains fixed and participates in every composed-scene rendering throughout the optimization process.

The fine-tuning objective is
\begin{equation}
    L_{\mathrm{train}}(t)
    =
    L_{\mathrm{rec}}(t)
    +
    \lambda_T L_{\mathrm{temp}}(t),
\end{equation}
where $L_{\mathrm{rec}}$ is the standard weighted combination of the $\ell_1$ and D-SSIM reconstruction terms evaluated on the complete rendered image. The temporal term applies only to the residual-sensitive subset.

We regularize frame-to-frame changes in scale, rotation, and opacity over the residual-sensitive subset. For normalized quaternions, $d_q(\mathbf q,\mathbf q')=\min(\|\mathbf q-\mathbf q'\|_2,\|\mathbf q+\mathbf q'\|_2)$ accounts for the equivalent representations $\mathbf q$ and $-\mathbf q$. The temporal term is
\begin{equation}
    \begin{aligned}
    L_{\mathrm{temp}}(t)
    ={}&
    \frac{\|\mathbf{s}_t^r-\mathbf{s}_{t-1}^r\|_F^2}{N_rD_{\mathbf{s}}}
    +
    \frac{1}{N_rD_{\mathbf{q}}}
    \sum_{i=1}^{N_r}
    d_q^2(\mathbf{q}_{t,i}^r,\mathbf{q}_{t-1,i}^r)\\
    &+
    \frac{\|\boldsymbol{\alpha}_t^r-\boldsymbol{\alpha}_{t-1}^r\|_F^2}{N_r},
    \end{aligned}
\end{equation}
where $N_r$ is the number of residual-sensitive primitives and $D_{\mathbf{s}}$ and $D_{\mathbf{q}}$ are the dimensions of scale and rotation. Residual translation already models positional change. Color is supervised through the composed reconstruction loss of the full scene.

Selective optimization keeps $G^{s}$ fixed within the GOP and produces frame-dependent attributes for $G^{r}_t$. The following subsection describes how the two representations are encoded as a reusable stable packet and a residual-sensitive Gaussian video.

\subsection{Dual-Stream Compression and Transmission}
\label{sec:compression}

After optimization, \textit{ResStreamGS} encodes the stable set $G^{s}$ once per GOP for reuse and represents the frame-dependent residual-sensitive sets $G^{r}_{t}$ as a compact Gaussian video stream.

\noindent\textbf{Stable Gaussian Packet.}
The attributes of $G^{s}$ remain fixed throughout the frame group $\mathcal{F}=\{t_0,\ldots,t_0+K-1\}$. The stable packet $\mathbf{B}^{s}$ is therefore encoded and transmitted once when the group is loaded. It stores the Gaussian attributes required for rendering together with the primitive indices needed to preserve their slots in the composed Gaussian set. The primitives are spatially ordered and packed by attribute type before entropy coding; the ordering, precision, and codec settings are reported in Sec.~\ref{sec:implementation_details}. Entropy coding is lossless on the quantized byte stream, while attribute quantization remains relative to the original floating-point representation. Keeping these persistent attributes outside the video path avoids repeated lossy coding and amortizes their transmission over $K$ frames.

\noindent\textbf{Residual-Sensitive Video Packaging.}
\textit{ResStreamGS} retains the Gaussian video interface of V$^3$~\cite{wang2024v3} but applies frame-wise H.264/AVC coding only to $G^{r}_{t}$, while $G^{s}$ remains outside the video path and is supplied by the reusable stable packet. This preserves hardware-decoder compatibility while restricting lossy temporal coding to primitives that require frame-dependent updates.

Each GOP is represented by $\mathcal{S}_{\mathcal{F}}=(\mathbf{B}^{s},\mathbf{V}^{r}_{t_0:t_0+K-1},\mathbf{M}_{\mathcal{F}})$. The metadata $\mathbf{M}_{\mathcal{F}}$ records primitive assignments, attribute layouts, quantization ranges, and synchronization information. It maps cached stable primitives and decoded residual-sensitive primitives back to their corresponding slots in the frame-wise Gaussian representation. The client loads $\mathbf{B}^{s}$ once per GOP and advances $\mathbf{V}^{r}$ frame by frame, reconstructing both subsets as a single primitive-level Gaussian scene rather than independently rendered sub-scenes.

\subsection{Mobile Playback Pipeline}
\label{sec:mobile}

At each GOP boundary, the client decompresses $\mathbf{B}^{s}$ and reconstructs $G^{s}$ using the metadata $\mathbf{M}_{\mathcal{F}}$ defined in Sec.~\ref{sec:compression}. The stable primitives are uploaded to a GPU-resident cache and reused throughout the group. For each frame, the platform hardware video decoder processes $\mathbf{V}^{r}$, from which the player reconstructs $G_t^{r}$. Only the residual-sensitive attributes and their corresponding GPU buffers are refreshed between frames. The cached and decoded primitives are composed before rendering:
\begin{equation}
    G_t
    =
    \Compose
    \left(
    G^{s},G_t^{r};\mathbf{M}_{\mathcal{F}}
    \right),
    \qquad
    \widehat{\mathbf I}_t
    =
    \Render(G_t).
\end{equation}
Composition is performed at the primitive level, allowing the two subsets to share visibility evaluation, depth ordering, and opacity accumulation within the same rendering pipeline instead of being rendered separately and blended in image space. The resulting Gaussian set then undergoes view-frustum culling and opacity pruning before depth sorting and Gaussian rasterization.

Keeping $G^{s}$ resident removes invariant attributes from the frame-level decode and buffer-update path, avoiding repeated stable-packet decompression, attribute reconstruction, and GPU upload. This caching alone does not reduce the number of primitives in $G_t$; without renderer-side filtering, the complete composed set still enters visibility processing. View-frustum culling rejects primitives outside the current view, while opacity pruning removes low-contribution primitives. These operations reduce the active Gaussian set passed to sorting and rasterization. Stable caching therefore avoids repeated decoding and state updates, whereas renderer-side filtering reduces sorting and rasterization workloads.

\section{Evaluation}
\label{sec:evaluation}

\begin{table*}[t]
  \centering
  \caption{Quantitative comparison of reconstruction quality and amortized streaming payload on Neu3D and MeetRoom. Neu3D results are averaged over five full-scene sequences. Mobile playback indicates whether a method supports on-device playback ($\checkmark$/$\times$), and payload denotes the complete decoder-required representation amortized per frame. Best and second-best results among compressed methods are shown in \textbf{bold} and \underline{underlined}, respectively.}
  \label{tab:main_results}
  \scriptsize
  \renewcommand{\arraystretch}{1.00}
  \begin{threeparttable}
  \begin{tabular}{@{}>{\centering\arraybackslash}p{0.20\textwidth}@{}>{\centering\arraybackslash}p{0.105\textwidth}@{}>{\centering\arraybackslash}p{0.11125\textwidth}@{}>{\centering\arraybackslash}p{0.11125\textwidth}@{}>{\centering\arraybackslash}p{0.11125\textwidth}@{}>{\centering\arraybackslash}p{0.11125\textwidth}@{}>{\centering\arraybackslash}p{0.15\textwidth}@{}}
    \toprule
    \textbf{Method} &
    \textbf{Mobile display} &
    \textbf{PSNR $\uparrow$} &
    \textbf{SSIM $\uparrow$} &
    \textbf{LPIPS $\downarrow$} &
    \textbf{DISTS $\downarrow$} &
    \textbf{Payload (MB) $\downarrow$} \\
    \midrule

    \multicolumn{7}{c}{\textbf{Neu3D}} \\
    \midrule\midrule
    Uncompressed Reference & -- & 34.60 & 0.9554 & 0.100 & 0.05188 & 107.810 \\
    4DGS & $\times$ & 29.70 & 0.9207 & \underline{0.121} & 0.06514 & \underline{0.895} \\
    GIFStream & $\times$ & \underline{30.59} & \underline{0.9275} & 0.151 & \underline{0.05658} & \textbf{0.039} \\
    V$^3$ & $\checkmark$ & 25.53 & 0.8276 & 0.320 & 0.16322 & 3.741 \\
    \rowcolor{ResStreamGreen}[0pt][0pt]
    \textbf{\textit{ResStreamGS}} & $\checkmark$ & \textbf{33.04} & \textbf{0.9425} & \textbf{0.107} & \textbf{0.03756} & 1.627 \\

    \midrule
    \multicolumn{7}{c}{\textbf{MeetRoom}} \\
    \midrule\midrule
    Uncompressed Reference & -- & 37.97 & 0.9621 & 0.046 & 0.05108 & 7.314 \\
    4DGS & $\times$ & \underline{31.97} & \underline{0.9569} & 0.166 & \textbf{0.07272} & 8.287 \\
    GIFStream & $\times$ & 31.02 & 0.9564 & \textbf{0.051} & 0.09013 & \textbf{0.032} \\
    V$^3$ & $\checkmark$ & 25.38 & 0.8239 & 0.277 & 0.23294 & 0.304 \\
    \rowcolor{ResStreamGreen}[0pt][0pt]
    \textbf{\textit{ResStreamGS}} & $\checkmark$ & \textbf{33.73} & \textbf{0.9606} & \underline{0.076} & \underline{0.09007} & \underline{0.239} \\
    \bottomrule
  \end{tabular}
  \end{threeparttable}
\end{table*}

\begin{figure*}[!t]
  \centering
    \includegraphics[width=0.90\textwidth]{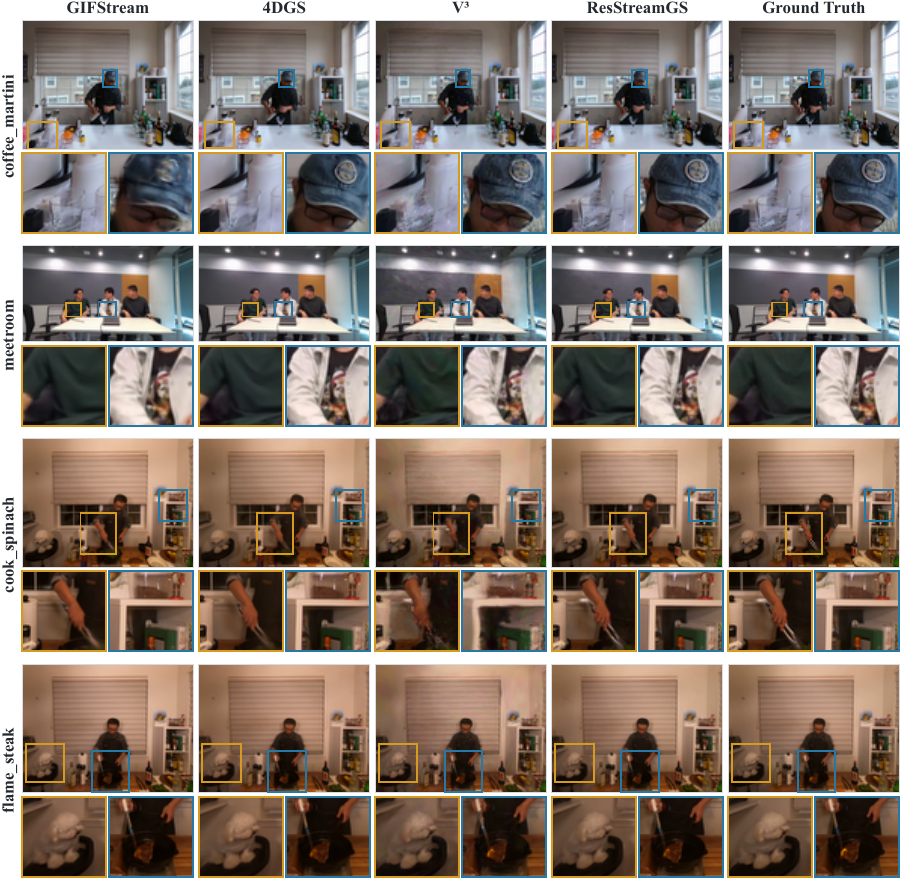}
  \caption{Qualitative comparison across representative scenes and methods.}
  \label{fig:qualitative_comparison_extended}
\end{figure*}

\subsection{Experimental Setup}
\label{sec:experimental_setup}

\noindent\textbf{Datasets and Evaluation Protocol.}
We evaluate \textit{ResStreamGS} on six complex full-scene sequences from two multiview video datasets, including five from Neu3D~\cite{li2022neural3dvideo} and one from MeetRoom~\cite{li2022streamrf}. These sequences contain localized dynamic content together with substantial room-scale structures, matching the streaming setting considered in this work. Each sequence contains 300 consecutive frames and is divided into 20-frame GOPs.

For each Neu3D sequence, we reserve five calibrated camera views for held-out evaluation and use the remaining views for training. MeetRoom contains 11 calibrated views, of which 10 are used for training and one is reserved for held-out evaluation. Neu3D rendering is evaluated at $1352\times1014$ resolution, while MeetRoom is evaluated at $640\times360$. The same frames, camera parameters, resolutions, GOP boundaries, training views, and held-out evaluation views are used for all locally evaluated methods. All rendering metrics are computed exclusively on held-out views across evaluated frames. Probe adaptation and quality-aware stream partitioning use only training views, and held-out views are not used for model optimization or stream-configuration selection. Neu3D results are averaged over all five evaluated sequences.

\noindent\textbf{Compared Methods.}
We use the complete per-frame Gaussian representations obtained after training but before quantization and stream coding as the trained-but-uncompressed reconstruction reference. We compare \textit{ResStreamGS} with 4DGS~\cite{wu2024d4gs}, GIFStream~\cite{li2025gifstream}, and a controlled V$^3$ All-Gaussians baseline~\cite{wang2024v3}. The V$^3$ baseline uses the same GOP structure, keyframe initialization, Gaussian-video attribute mapping, quantization scheme, codec, and QP configuration as \textit{ResStreamGS}. The difference lies in the temporal processing scope: the baseline keeps all Gaussian primitives in the frame-wise optimization and video-coding path, whereas \textit{ResStreamGS} transmits stable primitives once per GOP and applies frame-wise coding only to the residual-sensitive subset. Results obtained using method-specific protocols or reported by the original implementations are marked accordingly.

\noindent\textbf{Metrics and Payload Accounting.}
We evaluate reconstruction fidelity using PSNR, SSIM, AlexNet-based LPIPS, and DISTS~\cite{ding2022dists}. Metrics are computed from rendered images after complete stream decoding, Gaussian reconstruction and averaged over all evaluated frames and held-out camera views. Streaming efficiency is measured using the complete decoder-required representation amortized per frame. This payload includes all Gaussian data, video bitstreams, and metadata required to reconstruct and render each sequence. For ResStreamGS, the GOP-level stable packet and associated metadata are amortized over all frames within the corresponding GOP, while the frame-level residual-sensitive video stream is counted directly. The same deployment-oriented accounting principle is used for compared representations, while training-only artifacts such as checkpoints, optimizer states, and intermediate outputs are excluded.

\noindent\textbf{Mobile Testbed.}
Mobile playback is evaluated on an HONOR 50 Pro, a commercially available Android smartphone equipped with a Qualcomm Snapdragon 778G SoC and an Adreno 642L GPU. Our implementation uses standard Android hardware video decoding and GPU rendering interfaces without device-specific or vendor-specific tuning. We reimplement the V$^3$ Android player and verify that it reproduces the real-time ActorHQ playback performance reported in the original V$^3$ paper before using this implementation as our mobile baseline. Device-level evaluation covers all five Neu3D sequences using the complete 300-frame sequence for each scene. The V$^3$ baseline, the full \textit{ResStreamGS} system, and all mobile playback ablations are measured on the same device under identical rendering resolution, camera trajectory, and playback configuration.

\subsection{Implementation Details}
\label{sec:implementation_details}

All experiments use 20-frame GOPs, with the final frame of each group serving as the probe target. Probe adaptation is performed with Adam for 120 iterations. The first 20 iterations are treated as burn-in, and the remaining 100 states are used to estimate primitive-level trajectory statistics. We set $\lambda_{\mathrm{DSSIM}}=0.2$ and assign equal weights to the five attribute-specific sensitivity terms. Candidate residual-sensitive ratios from 0.5\% to 20\% are examined using a 1-percentage-point coarse search followed by 0.1-percentage-point local refinement. The composed-scene quality constraint is evaluated on the calibrated training views of the probe frame with $\delta=0.55$~dB. For each non-keyframe, residual translation and residual-sensitive fine-tuning are performed for 800 and 2,000 iterations, respectively, with $\lambda_T=10^{-3}$ for temporal regularization. The same optimization hyperparameters are used across all evaluated sequences and configurations unless otherwise stated.

Stable primitives are reordered in Morton order and serialized once per GOP. Positions and scales are stored in half precision, rotations as packed 32-bit quaternions, and color and opacity as 8-bit values before lossless compression with Zstandard. Residual-sensitive attributes are packed into 17 grayscale Gaussian-video planes. Position uses 16-bit quantization with separate low- and high-byte planes, while the remaining attributes use 8-bit quantization. The video planes are encoded with H.264/AVC using a base QP of 25. To preserve precision for more sensitive channels, position high-byte planes use QP~0, while color, scale, and rotation use QP~10. Position low-byte and opacity planes retain the base QP. The same coding configuration is used for all evaluated sequences.

\subsection{Rendering Quality and Streaming Efficiency}
\label{sec:quality_comparison}

Table~\ref{tab:main_results} summarizes the reconstruction quality and amortized payload of \textit{ResStreamGS} and the compared methods on Neu3D and MeetRoom. The controlled comparison with the V$^3$ All-Gaussians baseline isolates the effect of residual-sensitive separation under the same Gaussian-video optimization and coding framework. Across the two datasets, \textit{ResStreamGS} improves PSNR by 7.51--8.35~dB while reducing the amortized payload by 21.4--56.5\%. Rather than repeatedly optimizing and coding the complete Gaussian set, \textit{ResStreamGS} keeps temporally stable primitives outside the frame-wise video path and reserves per-frame updates for the residual-sensitive subset. These results show that matching the temporal granularity of processing to primitive behavior improves reconstruction fidelity while avoiding frame-wise coding.

\begin{figure}[t]
\centering
\vspace{0.33cm}
\makebox[\columnwidth][l]{%
\includegraphics[width=\dimexpr\columnwidth-6pt\relax]{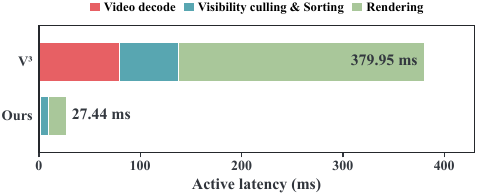}%
}
\caption{Client-side active latency comparison between V$^3$ and ResStreamGS during mobile playback. The total latency is decomposed into video decoding, visibility culling and sorting, and rendering.}
\label{fig:mobile_latency_comparison}
\end{figure}

\begin{figure}[t]
\centering
\makebox[\columnwidth][l]{%
\includegraphics[width=\dimexpr\columnwidth-8pt\relax]{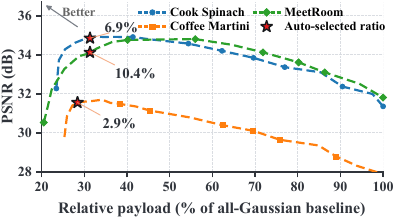}%
}
\caption{Quality--payload trade-off under different residual-sensitive Gaussian ratios on \textit{cook\_spinach}, \textit{coffee\_martini}, and MeetRoom. The ratio varies from 0.5\% to 100\%, with payload normalized by the all-Gaussians baseline. Stars denote the operating points automatically selected by the probe-stage quality criterion.}
\label{fig:ratio_payload_tradeoff}
\end{figure}

\begin{figure}[t]
\centering
\includegraphics[width=\columnwidth]{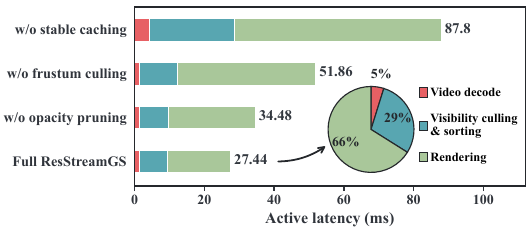}
\caption{Client-side active latency breakdown for ResStreamGS and its ablated variants during mobile playback. The ablations disable stable caching, view-frustum culling, and opacity pruning, respectively.}
\label{fig:latency_breakdown}
\end{figure}

The trained-but-uncompressed references in Table~\ref{tab:main_results} further indicate how much reconstruction quality is retained after stream packaging and decoding. On Neu3D, \textit{ResStreamGS} remains within 1.56~dB of the trained-but-uncompressed representation while reducing the amortized size from 107.810 to 1.627~MB/frame. The remaining gap reflects the combined effects of selective frame-wise optimization, attribute quantization, and lossy coding of the residual-sensitive stream. Nevertheless, the substantially smaller degradation than all-Gaussians video coding indicates that keeping stable primitives outside the repeated lossy coding path preserves much of the trained reconstruction fidelity.

The comparison with 4DGS and GIFStream provides a broader view of reconstruction quality and representation compactness. Among the compressed methods, \textit{ResStreamGS} achieves the highest PSNR and SSIM on both datasets and the lowest LPIPS on Neu3D, while GIFStream yields a smaller payload on both datasets and the lowest LPIPS on MeetRoom. 4DGS and GIFStream rely on neural attribute evaluation and are not designed around mobile hardware video decoding. Their compact representations therefore do not directly provide the hardware-decoder-oriented playback path targeted in this work. In contrast, \textit{ResStreamGS} retains an explicit Gaussian-video representation compatible with commodity hardware codecs while achieving higher reconstruction fidelity on the evaluated datasets. The qualitative comparisons in Fig.~\ref{fig:qualitative_comparison_extended} show the same trend, with \textit{ResStreamGS} preserving local structures and dynamic details more faithfully. The reported payload measures the complete decoder-required representation amortized per frame and should therefore be interpreted as representation-level streaming cost rather than end-to-end wireless network performance.

The representation-level improvements also translate into lower client-side processing cost during mobile playback. Figure~\ref{fig:mobile_latency_comparison} shows that active latency decreases from 379.95 to 27.44~ms, while the measured playback rate increases from 2.614 to 30.883~FPS and memory usage decreases from 518.4 to 331.9~MiB. These gains arise from complementary mechanisms: stable reuse removes persistent Gaussian attributes from repeated frame-wise decoding, reconstruction, and GPU buffer updates, while renderer-side filtering reduces the active primitive set reaching sorting and rasterization. The reduction in downstream rendering workload should therefore not be attributed to stable caching alone. Their individual contributions are examined separately in Sec.~\ref{sec:mobile_analysis}. Active latency reports the accumulated execution time of the profiled client-side stages, whereas FPS reflects the complete playback loop and additionally includes synchronization and system-level overheads; the two values are therefore not direct reciprocals.

\subsection{Ablation Study}
\label{sec:ablation}

We first examine two complementary decisions in residual-sensitive partitioning: which primitives should remain frame-wise adaptive, and how large that subset should be. To isolate the first question, Table~\ref{tab:selection_ablation} fixes the dynamic ratio at 8.00\% for all ranking strategies. Under this matched budget, the all-attribute ranking improves PSNR by 2.798~dB over Random Selection while maintaining a comparable payload. This result shows that reconstruction quality depends strongly on which primitives are retained for frame-wise adaptation rather than simply on their number. Position-only ranking reaches 31.193~dB, indicating that spatial displacement provides a strong first-order cue for temporal variation. Adding rotation, scale, opacity, or color individually yields gains of 0.380, 0.194, 0.144, and 0.129~dB, respectively. In contrast, combining all five attributes improves PSNR to 32.745~dB, a 1.552-dB gain over position alone with only a minor payload difference. The substantially larger joint gain indicates that these cues provide complementary information: primitives with limited positional displacement can still require frame-wise updates because changes in orientation, scale, opacity, or appearance alter their projected contribution. Multi-attribute profiling therefore improves the ranking itself rather than relying on a larger dynamic budget.

With the ranking fixed, varying the dynamic ratio addresses the complementary question of how much frame-wise adaptability is useful. Figure~\ref{fig:ratio_payload_tradeoff} reveals a consistent three-stage behavior across the evaluated scenes. At small ratios, reconstruction quality improves rapidly as the highest-ranked residual-sensitive primitives are added, indicating that an insufficient dynamic subset cannot capture important temporal variations. After these primitives have been included, the gain gradually saturates, suggesting that most useful frame-dependent changes are concentrated in a relatively small fraction of the Gaussian set. Increasing the ratio further provides little additional benefit and can even reduce reconstruction quality while continuously increasing payload. This high-ratio behavior is consistent with the failure mode identified in Sec.~\ref{sec:motivation}: increasingly stable primitives are brought back into frame-wise optimization and lossy video coding, where they are again exposed to optimization drift and codec distortion despite requiring little temporal adaptation. In the limiting case, a fully dynamic partition approaches the all-Gaussians processing paradigm that \textit{ResStreamGS} is designed to avoid. The desirable operating point is therefore a balance between sufficient temporal adaptability and stable reuse.

The automatically selected operating points follow this trade-off while adapting to scene-specific temporal structure. Their dynamic ratios range from 2.9\% to 10.4\%, rather than converging to a common fixed proportion. On both evaluated Neu3D sequences, the selected configurations remain within 0.15~dB of the best sampled post-codec PSNR while requiring less payload. MeetRoom exhibits a different trade-off: the selected 10.4\% configuration is 0.68~dB below the best sampled 40\% configuration but reduces payload by 44.1\%. This does not indicate that the selector fails to recover the post-codec PSNR maximum. The selection criterion instead searches for the smallest residual-sensitive subset satisfying the probe-stage composed-scene quality constraint before video coding, and therefore favors a compact operating point once sufficient quality is preserved. A single fixed ratio would either retain unnecessary frame-wise updates in some scenes or remove useful adaptability in others. Together, multi-attribute profiling determines which primitives should remain adaptive, while quality-aware selection determines how many are necessary.

\begin{table}[t]
  \centering
    \caption{Ablation of Gaussian ranking cues on \textit{coffee\_martini}.}
  \label{tab:selection_ablation}
  \scriptsize
  \renewcommand{\arraystretch}{1.10}
  \begin{tabular}{@{}>{\raggedright\arraybackslash}p{0.25\columnwidth}
  !{\color{black}\vrule width\lightrulewidth}
  >{\centering\arraybackslash}p{0.20\columnwidth}
  @{\hspace{2.8pt}}
  >{\centering\arraybackslash}p{0.33\columnwidth}@{}}
    \toprule
    \hspace{1.2em}\textbf{Ranking score} & \textbf{PSNR (dB) $\uparrow$} & \textbf{Payload (MB/frame) $\downarrow$} \\
    \midrule
    \hspace{2.2em}Random & 29.947 & 1.805 \\
    \midrule
    \hspace{2.2em}Position & 31.193 & 1.846 \\
    \hspace{2.6em}+ Rotation & 31.573 & 1.861 \\
    \hspace{2.6em}+ Scale & 31.387 & 1.848 \\
    \hspace{2.6em}+ Opacity & 31.337 & 1.844 \\
    \hspace{2.6em}+ Color & 31.322 & 1.859 \\
    \midrule
    \hspace{1.2em}\textbf{All attributes} & \textbf{32.745} & \textbf{1.853} \\
    \bottomrule
  \end{tabular}
\end{table}

Finally, we examine whether using the terminal frame as the probe target provides a representative partition for other frames within a GOP. For every frame in the evaluated 300-frame sequences, we measure the post-codec PSNR gap relative to its corresponding trained-but-uncompressed representation and group the results by temporal distance from the probe frame. The quality gap does not increase with distance: the fitted slopes are $-0.021$, $-0.020$, and $-0.105$~dB per frame for \textit{cook\_spinach}, \textit{coffee\_martini}, and MeetRoom, respectively. We further examine the 20\% of frames with the largest inter-frame appearance changes, whose average gaps differ from the corresponding scene averages by only $+0.37$, $-0.05$, and $-0.50$~dB. These observations indicate that, under the 20-frame GOP setting used in our experiments, the terminal-frame probe remains informative for constructing a shared GOP-level partition even when intermediate frames exhibit pronounced temporal variation. This further supports the practical use of a single lightweight probe per GOP, while longer GOPs or scenes with abrupt structural changes may require more adaptive partitioning.

\subsection{Mobile Playback Analysis}
\label{sec:mobile_analysis}

We examine how the client-side components of \textit{ResStreamGS} contribute to mobile playback performance in Table~\ref{tab:mobile_performance}. Stable caching has the largest effect on playback throughput. Removing it increases active latency from 27.44 to 87.80~ms and reduces the measured frame rate from 30.883 to 10.392~FPS. Without caching, stable Gaussian attributes must be reconstructed and uploaded rather than remaining resident across the GOP. Stable caching therefore converts temporal persistence into persistent device state and avoids repeated preparation of primitives without frame-wise updates.

\begin{table}[t]
  \centering
  \caption{Mobile playback performance and component ablations.}
  \label{tab:mobile_performance}
  \footnotesize
  \setlength{\tabcolsep}{2.35pt}
  \renewcommand{\arraystretch}{1.13}
  \begin{tabular}{l!{\color{black}\vrule width\lightrulewidth}ccc}
    \toprule
    \textbf{Configuration} & \textbf{Latency (ms)$\downarrow$} & \textbf{FPS$\uparrow$} & \textbf{Memory (MiB)$\downarrow$} \\
    \midrule
    V$^3$ & 379.95 & 2.614 & 518.4 \\
    w/o stable caching & 87.80 & 10.392 & 470.4 \\
    w/o frustum culling & 51.86 & 16.859 & 656.2 \\
    w/o opacity pruning & 34.48 & 24.531 & 325.0 \\
    \midrule
    \textbf{Full ResStreamGS} & \textbf{27.44} & \textbf{30.883} & \textbf{331.9} \\
    \bottomrule
  \end{tabular}
\end{table}

Renderer-side filtering reduces a different part of the playback workload. Disabling view-frustum culling increases active latency from 27.44 to 51.86~ms and memory usage from 331.9 to 656.2~MiB because off-view primitives remain active. Removing opacity pruning increases latency to 34.48~ms and reduces playback to 24.531~FPS, while memory remains comparable. The three mechanisms therefore play distinct roles: stable caching avoids repeated reconstruction and GPU updates, frustum culling removes viewpoint-irrelevant primitives, and opacity pruning removes low-contribution primitives before sorting and rendering.

The stage-level breakdown in Fig.~\ref{fig:latency_breakdown} further supports this interpretation. Removing stable caching produces the largest increase because persistent Gaussian state re-enters the frame-wise preparation path. Disabling frustum culling mainly increases downstream sorting and rendering workloads, whereas disabling opacity pruning causes a smaller increase near the end of the pipeline. Since these mechanisms act at different stages, their benefits are complementary rather than additive. These ablations reflect component-wise effects within \textit{ResStreamGS}, rather than a strict decomposition of the total gap between V$^3$ and the full system. The reduction from 379.95 to 27.44~ms results from the combined dual-stream representation, stable reuse, and renderer-side filtering, which together sustain real-time playback on the evaluated device.

\section{Conclusion}
\label{sec:discussion}

In this paper, we present ResStreamGS, a residual-sensitive dual-stream framework for mobile volumetric video streaming. The method separates Gaussian primitives according to their temporal behavior, reusing stable Gaussians at the GOP level while retaining residual-sensitive Gaussians in the frame-wise video stream. To determine this partition, ResStreamGS employs lightweight probe adaptation, multi-attribute residual-sensitivity profiling, and quality-aware ratio selection. Extensive experiments demonstrate that ResStreamGS improves post-codec reconstruction quality while reducing streaming payload, achieving 7.51--8.35~dB higher PSNR than the controlled V$^3$-style baseline, up to 56.5\% payload reduction, and 30.883~FPS on commodity mobile hardware.

Although ResStreamGS shows substantial improvements, efficient full-scene volumetric video streaming remains challenging. Our method uses a fixed Gaussian partition within each GOP, which may not capture large or abrupt temporal variations. In addition, the residual-sensitive ratio is selected before video coding and does not explicitly account for codec distortion or bitrate. Future work will explore adaptive GOP partitioning, codec-aware ratio selection, and more flexible temporal modeling to further improve streaming quality and efficiency under diverse scene dynamics.

\clearpage
\bibliographystyle{abbrv-doi}

\bibliography{template}
\end{document}